# GARCH modelling in continuous time for irregularly spaced time series data

ROSS A. MALLER[1], GERNOT MÜLLER[2] and ALEX SZIMAYER[3]

[1]*School of Finance & Applied Statistics and Center for Mathematics & Its Applications, The Australian National University, ACT 0200, Australia. E-mail: Ross.Maller@anu.edu.au*
[2]*Center for Mathematical Sciences, Munich University of Technology, Boltzmannstraße 3, 85747 Garching, Germany. E-mail: mueller@ma.tum.de*
[3]*Financial Mathematics, Fraunhofer ITWM, Fraunhofer-Platz 1, 67663 Kaiserslautern, Germany. E-mail: alexander.szimayer@itwm.fraunhofer.de*

The discrete-time GARCH methodology which has had such a profound influence on the modelling of heteroscedasticity in time series is intuitively well motivated in capturing many 'stylized facts' concerning financial series, and is now almost routinely used in a wide range of situations, often including some where the data are not observed at equally spaced intervals of time. However, such data is more appropriately analyzed with a continuous-time model which preserves the essential features of the successful GARCH paradigm. One possible such extension is the diffusion limit of Nelson, but this is problematic in that the discrete-time GARCH model and its continuous-time diffusion limit are not statistically equivalent. As an alternative, Klüppelberg *et al.* recently introduced a continuous-time version of the GARCH (the 'COGARCH' process) which is constructed directly from a background driving Lévy process. The present paper shows how to fit this model to irregularly spaced time series data using discrete-time GARCH methodology, by approximating the COGARCH with an embedded sequence of discrete-time GARCH series which converges to the continuous-time model in a strong sense (in probability, in the Skorokhod metric), as the discrete approximating grid grows finer. This property is also especially useful in certain other applications, such as options pricing. The way is then open to using, for the COGARCH, similar statistical techniques to those already worked out for GARCH models and to illustrate this, an empirical investigation using stock index data is carried out.

## 1. Introduction

The modelling of time series in finance, economics and other fields frequently has to account for heteroscedasticity in the underlying data. Popular approaches to this problem use the autoregressive conditional heteroscedasticity (ARCH) model of Engle [6] and its generalized version, the GARCH model of Bollerslev [4]. The main principle of time series modelling using GARCH is that a 'large' innovation (or unexpected change) in a period increases the variance of the innovation in the following periods. This constitutes a feedback mechanism whereby a single univariate series of innovations determines both the time series and its conditional variance structure.

The GARCH concept has had a profound influence on time series modelling. Many other stochastic volatility models have been proposed, but the GARCH remains one of the easiest to conceptualize, is well established and thoroughly studied from a theoretical point of view and has been successfully applied in many practical situations. Some measure of the volatility (or risk) of an asset price is crucial in a wide variety of risk management areas (e.g., Jorion [11], Chapter 8.2, page 186, McNeil and Frey [19]) and in the valuation of financial derivatives (e.g., Ritchken and Trevor [23]).

In practice, for various reasons, including weekend and holiday effects, or in tick-by-tick data, many financial time series are irregularly spaced and this, together with options pricing requirements, in particular, has created a demand for continuous-time models. Nelson [21] suggested that GARCH models be seen as discrete approximations to diffusions. He showed that some standard GARCH models, when scaled in certain ways on an approximating grid, converge in distribution, as the grid grows finer, to a bivariate diffusion process, the variance rate (or volatility) of which exhibits mean reverting behavior. Nelson's result served for some time as a justification for statistical inference of continuous-time models using GARCH as an approximation.

However, in Nelson's setup, the limiting process involves two independent Brownian motions, one of which drives the volatility and the other the accumulated time series (which then becomes a stochastic integral). This runs quite counter to the philosophy of the original GARCH paradigm, whereby a single univariate series of innovations drives both mean and variance equations, thus providing a feedback mechanism. It is possible to modify Nelson's diffusion approximation so as to obtain convergence in distribution to a process which is driven by a single Brownian motion; however, the limit then has a deterministic volatility and the GARCH features disappear (see Corradi [5]). As a further problematic aspect, Wang [26] showed that a GARCH model and its continuous-time diffusion limit are not statistically equivalent, except in the case of the deterministic volatility limit derived by Corradi. This means that parameter estimation and testing for an underlying continuous-time diffusion model with stochastic volatility cannot be accomplished using a GARCH approximation in discrete-time.

Recently, Klüppelberg, Lindner and Maller [14] introduced a continuous-time version of the GARCH model, which they dubbed the 'COGARCH'. In contrast to the approaches of Nelson and Corradi based on limiting diffusions, [14] starts with a pure jump Lévy process and generalizes a discrete recursion which lies at the heart of the GARCH. In this way, the main characteristics of the original GARCH are preserved: a single univariate process drives both the volatility process and the integrated GARCH process itself, and the same sort of feedback mechanism is built into the continuous-time model, that is, a large change in the Lévy process results in an increase of the volatility, as well as simultaneously increasing or decreasing the level of the process.

In the present paper, we study approximations to a COGARCH and estimation of its parameters when the COGARCH is the underlying data generating process. We show how the COGARCH can be obtained as the limit of an embedded sequence of discrete-time GARCH series. This demonstrates that Nelson's bivariate diffusion limit is not the only possible limit of a sequence of GARCH models and is, perhaps, not even the most natural. Further, our approach suggests how statistical techniques developed for GARCH models can be carried over to the COGARCH, after appropriate rescaling, to match the discrete- and continuous-time parameter sets. This allows us, in particular, to overcome difficulties associated with the analysis of irregularly spaced data. To illustrate, we carry out an empirical investigation using ASX200 stock index data, and some simulations. The convergence of the discrete- to the continuous-time processes is shown to be in probability in the Skorokhod metric and is therefore stronger than the previously mentioned weak convergence results of Nelson and Corradi.

While there are studies on discretely observed diffusions (see, e.g., [1] and [10] for recent references), very little has been done with jump processes in our context. But, recently, Kallsen and Vesenmayer [13] have obtained COGARCH as a *weak* limit of embedded GARCH series (see also [12]). Their approach is quite different to ours, proceeding by way of the infinitesimal generator of the bivariate Markov process representation of the COGARCH process. In our setup, the embedded GARCH models and the COGARCH model are defined on the same probability space and pathwise arguments are invoked when proving the convergence. There are areas of applications where the stronger convergence is essential, for example, in the pricing of American options (see [17] and their discussion and references).

In related investigations, Müller [20] developed a Markov chain Monte Carlo estimation procedure for the parameters of a COGARCH, which is applicable to irregularly spaced data. However, it assumes quite detailed knowledge about the driving Lévy process and is heavily computer intensive, so simulations using it are currently infeasible. Haug *et al.* [9] use a method of moments procedure for COGARCH parameter estimation, but this is not easily adapted for unequally spaced series.

Our paper is organized as follows. Section 2 briefly recalls the GARCH and COGARCH models and the main convergence result, Theorem 2.1, is stated. In Section 3, an estimation procedure for the COGARCH parameters is proposed, applied to a financial data set and supported by a Monte Carlo study. In Section 4, we discuss the implications of our results, especially with reference to Wang's [26] far reaching observation. All proofs are contained in Section 5.

## 2. Setup and convergence theorem

To begin, we recall the definition of a continuous-time GARCH process, as introduced in [14]. On a filtered probability space $(\Omega, \mathcal{F}, \mathbb{P}, (\mathcal{F}_t)_{t\geq 0})$ satisfying the 'usual hypothesis' (see Protter, [22], page 3), we are given a *background driving Lévy process* $L = (L(t))_{t\geq 0}$, that is, a real-valued, pure jump Lévy process with characteristic triplet $(\gamma, 0, \Pi)$ and $L(0) = 0$. Thus, it has characteristic function satisfying

$$\mathbb{E}\mathrm{e}^{\mathrm{i}\theta L(t)} = \exp\biggl(\mathrm{i}t\gamma\theta + t\int_{\mathbb{R}\setminus\{0\}}(\mathrm{e}^{\mathrm{i}\theta x} - 1 - \mathrm{i}\theta x\mathbf{1}_{\{|x|\leq 1\}})\Pi(\mathrm{d}x)\biggr), \qquad t \geq 0;$$

see [2, 3] and [24] for detailed background and results concerning Lévy processes. The Lévy measure $\Pi$ is a measure on the Borel subsets of $\mathbb{R} \setminus \{0\}$, and $\gamma$ is a constant depending on the truncation at 0; we choose the standard truncation $\mathbf{1}_{\{|x| \leq 1\}}$. The filtration $(\mathcal{F}_t)_{t \geq 0}$ is the completed natural filtration of the Lévy process $L$. Note that no Brownian component is present in the Lévy process; we show later how it can be included if desired. We suppose throughout that $\mathbb{E}L(1) = 0$ and $\mathbb{E}L^2(1) = 1$.

Given parameters $(\beta, \eta, \varphi)$, with $\beta > 0$, $\eta > 0$, $\varphi \geq 0$, and a square-integrable random variable (r.v.) $\sigma(0)$ independent of $L$, the *COGARCH variance process* $\sigma^2 = (\sigma^2(t))_{t \geq 0}$ is defined as the almost surely (a.s.) unique solution of the stochastic differential equation (SDE)

$$\mathrm{d}\sigma^2(t) = (\beta - \eta \sigma^2(t-))\,\mathrm{d}t + \varphi \sigma^2(t-)\,\mathrm{d}[L, L](t), \qquad t > 0, \tag{2.1}$$

where $[L, L]$ is the bracket process (quadratic variation) of $L$ (Protter [22], page 66). We then define the *integrated COGARCH process* $G = (G(t))_{t \geq 0}$ in terms of $L$ and $\sigma$ as

$$G(t) = \int_0^t \sigma(s-)\,\mathrm{d}L(s), \qquad t \geq 0. \tag{2.2}$$

We refer to [14] and [15] for detailed properties of $G$ and $\sigma^2$.

### 2.1. Approximating the COGARCH

Our aim is to define a family of discrete-time processes, $G_n = (G_n(t))_{t \geq 0}$, $n = 1, 2, \ldots$, constructed from a GARCH$(1, 1)$ process, which approximates the continuous-time process $G$. This allows us to take advantage of widely used inferential and other methods in time series modelling and econometrics for this well-established process class. After appropriate rescaling to match the discrete- and continuous-time parameter sets, $G_n$ will be shown to converge to $G$ in a quite strong sense.

The discretization is over a finite interval $[0, T]$, $T > 0$, and is operationalized as follows. Take deterministic sequences $(N_n)_{n \geq 1}$ with $\lim_{n \to \infty} N_n = \infty$ and $0 = t_0(n) < t_1(n) < \cdots < t_{N_n}(n) = T$, and, for each $n = 1, 2, \ldots$, divide $[0, T]$ into $N_n$ subintervals of length $\Delta t_i(n) := t_i(n) - t_{i-1}(n)$ for $i = 1, 2, \ldots, N_n$. Assume $\Delta t(n) := \max_{i=1,\ldots,N_n} \Delta t_i(n) \to 0$ as $n \to \infty$ and define, for each $n = 1, 2, \ldots$, a discrete-time process $(G_{i,n})_{i=1,\ldots,N_n}$ satisfying

$$G_{i,n} = G_{i-1,n} + \sigma_{i-1,n}\sqrt{\Delta t_i(n)}\varepsilon_{i,n}, \qquad i = 1, 2, \ldots, N_n, \tag{2.3}$$

where $G_{0,n} = G(0) = 0$ and the variance $\sigma_{i,n}^2$ follows the recursion

$$\sigma_{i,n}^2 = \beta \Delta t_i(n) + (1 + \varphi \Delta t_i(n) \varepsilon_{i,n}^2) \mathrm{e}^{-\eta \Delta t_i(n)} \sigma_{i-1,n}^2, \qquad i = 1, 2, \ldots, N_n. \tag{2.4}$$

Here, the innovations $(\varepsilon_{i,n})_{i=1,\ldots,N_n}$, $n = 1, 2, \ldots$, are constructed using a 'first jump' approximation to the Lévy process, as follows. Take a strictly positive sequence $1 \geq m_n \downarrow 0$ of reals satisfying $\lim_{n \to \infty} \Delta t(n) \overline{\Pi}^2(m_n) = 0$, where $\overline{\Pi}(x) = \int_{|y|>x} \Pi(\mathrm{d}y)$ is the tail of $\Pi$. Such a sequence always exists, as $\lim_{x\downarrow 0} x^2\overline{\Pi}(x) = 0$ for any Lévy measure. Let $\Delta L(t) = L(t) - L(t-)$, $t > 0$, $\Delta L(0) = 0$. Fix $n \geq 1$ and define stopping times $\tau_{i,n}$ by

$$\tau_{i,n} = \inf\{t \in [t_{i-1}(n), t_i(n)) : |\Delta L(t)| \geq m_n\}, \qquad i = 1, \ldots, N_n. \tag{2.5}$$

(Throughout, an infimum over the empty set is understood as being $+\infty$.) $\tau_{i,n}$ is the time of the first jump of $L$ in the $i$th interval whose magnitude exceeds $m_n$, if such a jump occurs.

By the strong Markov property, $(\mathbf{1}_{\{\tau_{i,n}<\infty\}}\Delta L(\tau_{i,n}))_{i=1,\ldots,N_n}$ is for each $n = 1, 2, \ldots$ a sequence of independent r.v.s, with distribution specified by:

$$\frac{\Pi(\mathrm{d}x)\mathbf{1}_{\{|x|>m_n\}}}{\overline{\Pi}(m_n)}(1 - \mathrm{e}^{-\Delta t_i(n)\overline{\Pi}(m_n)}), \qquad x \in \mathbb{R}\setminus\{0\}, i = 1, 2, \ldots, N_n, \tag{2.6}$$

and with mass $\mathrm{e}^{-\Delta t_i(n)\overline{\Pi}(m_n)}$ at 0. These r.v.s have finite mean, $\nu_i(n)$, and variance, $\xi_i(n)$, say, since $\mathbb{E}L^2(1)$ is finite. The innovations series $(\varepsilon_{i,n})_{i=1,\ldots,N_n}$ required for (2.3) is now defined by

$$\varepsilon_{i,n} = \frac{\mathbf{1}_{\{\tau_{i,n}<\infty\}}\Delta L(\tau_{i,n}) - \nu_i(n)}{\xi_i(n)}, \qquad i = 1, 2, \ldots, N_n. \tag{2.7}$$

For each $n = 1, 2, \ldots$, the $\varepsilon_{i,n}$ are independent with $\mathbb{E}\varepsilon_{1,n} = 0$ and $\text{Var}(\varepsilon_{1,n}) = 1$. Finally, in (2.4), we take $\sigma_{0,n}^2 = \sigma^2(0)$, independent of the $\varepsilon_{i,n}$.

*Remark 2.1.* Equations (2.3) and (2.4) specify a GARCH(1,1)-type recursion in the following sense. In the ordinary discrete-time GARCH(1,1) series, the volatility sequence satisfies

$$\sigma_i^2 = a + b\sigma_{i-1}^2\varepsilon_{i-1}^2 + c\sigma_{i-1}^2 \tag{2.8}$$

for constants $a$, $b$, $c$. When the time grid is equally spaced so that $\Delta t_i(n) = \Delta t(n)$, $i = 1, 2, \ldots, N_n$, (2.4) is equivalent to (2.8), after rescaling by $\Delta t(n)$ and a reparametrization from $(\beta, \varphi, \eta)$ to $(a, b, c)$, and (2.3) becomes a rescaled GARCH equation for the differenced sequence $G_{i,n} - G_{i-1,n}$. More generally, with an unequally spaced grid, if the series are scaled as in (2.3) and (2.4), convergence to the COGARCH is obtained, as we show next.

Embed the discrete-time processes $G_{\cdot,n}$ and $\sigma_{\cdot,n}^2$ into continuous-time versions $G_n$ and $\sigma_n^2$ defined by

$$G_n(t) := G_{i,n} \quad \text{and} \quad \sigma_n^2(t) := \sigma_{i,n}^2, \qquad \text{when } t \in [t_{i-1}(n), t_i(n)), 0 \leq t \leq T, \tag{2.9}$$

with $G_n(0) = 0$. The processes $G_n$ and $\sigma_n$ are in $\mathbb{D}[0,T]$, the space of càdlàg real-valued stochastic processes on $[0,T]$. Recall that the Skorokhod $J_1$ distance between two, $\mathbb{R}^d$-valued processes $U, V$, each in $\mathbb{D}^d[0,T]$ (the space of càdlàg $\mathbb{R}^d$-valued stochastic processes on $[0, T])$, is defined by

$$\rho(U, V) = \inf_{\lambda \in \Lambda} \left\{ \sup_{0 \le t \le T} \|U_t - V_{\lambda(t)}\| + \sup_{0 \le t \le T} |\lambda(t) - t| \right\}, \tag{2.10}$$

where $\Lambda$ is the set of strictly increasing continuous functions with $\lambda(0) = 0$ and $\lambda(T) = T$ (Gihman and Skorokhod [7], page 470). We can now state our main result for this section.

**Theorem 2.1.** *In the above setup, the Skorokhod distance between the processes $(G, \sigma^2)$ defined by (2.1) and (2.2), and the discretized, piecewise constant processes $(G_n, \sigma_n^2)_{n \ge 1}$ defined by (2.9), converges in probability to 0 as $n \to \infty$, that is,*

$$\rho((G_n, \sigma_n^2), (G, \sigma^2)) \xrightarrow{\mathbb{P}} 0 \qquad \text{as } n \to \infty. \tag{2.11}$$

*Consequently, we also have convergence in distribution in $\mathbb{D}[0, T] \times \mathbb{D}[0, T]$: $(G_n, \sigma_n^2) \xrightarrow{\mathbb{D}} (G, \sigma^2)$ as $n \to \infty$.*

*Remark 2.2.* (i) The derivation in [14] of the COGARCH employs an auxiliary Lévy process $X = (X(t))_{t \ge 0}$ constructed from $L$, $\eta > 0$ and $\varphi \ge 0$:

$$X(t) = \eta t - \sum_{0 < s \le t} \log(1 + \varphi(\Delta L(s))^2), \qquad t \ge 0. \tag{2.12}$$

$X$ is a spectrally negative Lévy process of bounded variation. (In (2.12), we have adopted the parameterization of [9], which differs somewhat from that of [14]; the latter used $0 < \delta < 1$ whereas we use $\eta > 0$, with $e^{-\eta} = \delta$, and used $\lambda/\delta$, for another parameter $\lambda \ge 0$, whereas we use $\varphi$.) Using Itô's lemma, we can verify that the solution to (2.1) can be written in terms of $X$ as

$$\sigma^2(t) = \left( \beta \int_0^t e^{X(s)} \, ds + \sigma^2(0) \right) e^{-X(t)}, \qquad t \ge 0. \tag{2.13}$$

This shows $\sigma^2(t)$ to be a kind of generalized Ornstein–Uhlenbeck (OU) process (cf. [18]), parameterized by $(\beta, \eta, \varphi)$ and driven by the process $L$.

(ii) Our procedure can be generalized to include a Brownian component. Let $B = (B(t))_{t \ge 0}$ be a standard Brownian motion and $L$ an independent, pure jump Lévy process with finite variance, and define $L^\dagger = \varsigma B + L$, where $\varsigma > 0$. Using $L^\dagger$ in place of $L$ in (2.1) and (2.2) introduces a diffusion component into the COGARCH. Center and scale so that $\mathbb{E}L^\dagger(1) = 0$ and $\mathbb{E}(L^\dagger(1))^2 = \varsigma^2 + \int x^2 \Pi(dx) = 1$. The convergence result of Theorem 2.1 extends to this setting if we modify the definition of the process $X$ in (2.12) to

$$X^\dagger(t) = (\eta - \varphi \varsigma^2) t - \sum_{0 < s \le t} \log(1 + \varphi(\Delta L^\dagger(s))^2), \qquad t \ge 0.$$

The term $\varphi \varsigma^2$ results from the bracket process of $B$. For a related convergence result, see Theorem 2.2 of Szimayer and Maller [25].

(iii) Now suppose that the modified COGARCH in the previous remark is, in fact, driven by a pure diffusion, that is, $L^\dagger = B$. The COGARCH then reduces to the process obtained in the limit by Corradi [5] and the GARCH approximations converge to Corradi's deterministic volatility limit. In this simplified situation, the GARCH models and the diffusion limit are statistically equivalent, as shown by Wang [26].

(iv) We have restricted ourselves throughout to convergence on the compact interval $[0, T]$. This is true for every $T > 0$, although the approximating processes depend on $T$ in a non-essential way. It is not difficult to modify our setup slightly so as to get approximating processes which converge to $(G, \sigma^2)$ uniformly on compacts (u.c.p., in the terminology of [22], page 57) and, consequently, also in $\mathbb{D}[0, \infty) \times \mathbb{D}[0, \infty)$ [16]. We omit the details here.

Theorem 2.1 is proved in Section 5. Next, we illustrate how to use the convergence result to analyze irregularly spaced time series data.

## 3. GARCH analysis of irregularly spaced data

In this section, we apply the insights gained by our discrete approximation of the continuous-time GARCH process to suggest a method of fitting the model to unequally spaced times series data. We build on the well-understood methodology developed for the discrete-time GARCH.

Suppose we have observations $G(t_i)$, $0 = t_0 < t_1 < \cdots < t_N = T$, on the integrated COGARCH as defined and parameterized in (2.1) and (2.2), assumed to be in its stationary regime. The $\{t_i\}$ are assumed fixed (non-random) time points. Let $Y_i = G(t_i) - G(t_{i-1})$ denote the observed returns and let $\Delta t_i := t_i - t_{i-1}$. From (2.2), we can then write

$$Y_i = \int_{t_{i-1}}^{t_i} \sigma(s-)\,\mathrm{d}L(s), \tag{3.1}$$

where $L$ is a Lévy process with $\mathbb{E}L(1) = 0$ and $\mathbb{E}L^2(1) = 1$ assumed.

Our aim is to use a pseudo-maximum likelihood (PML) method to estimate the parameters $(\beta, \eta, \varphi)$ from the observed $Y_1, Y_2, \ldots, Y_N$. To derive the pseudo-likelihood function, observe that, because $\sigma$ is Markovian ([14], Theorem 3.2), $Y_i$ is conditionally independent of $Y_{i-1}, Y_{i-2}, \ldots$, given $\mathcal{F}_{t_{i-1}}$. We have $\mathbb{E}(Y_i | \mathcal{F}_{t_{i-1}}) = 0$ for the conditional expectation of $Y_i$, and, for the conditional variance,

$$\rho_i^2 := \mathbb{E}(Y_i^2 | \mathcal{F}_{t_{i-1}}) = \left(\sigma^2(t_{i-1}) - \frac{\beta}{\eta - \varphi}\right)\left(\frac{\mathrm{e}^{(\eta-\varphi)\Delta t_i} - 1}{\eta - \varphi}\right) + \frac{\beta \Delta t_i}{\eta - \varphi}. \tag{3.2}$$

Equation (3.2) follows from the calculation in the third display on page 618 of [14]. To ensure stationarity, we take $\mathbb{E}\sigma^2(0) = \beta/(\eta - \varphi)$, with $\eta > \varphi$, in that formula and, in our setting, $\int_\mathbb{R} y^2 \Pi(\mathrm{d}y) = \mathbb{E}L^2(1) = 1$.

Applying the PML method, then, we assume the $Y_i$ are conditionally $N(0, \rho_i^2)$ and use recursive conditioning to write a pseudo-log-likelihood function for $Y_1, Y_2, \ldots, Y_N$ as

$$\mathcal{L}_N = \mathcal{L}_N(\beta, \varphi, \eta) = -\frac{1}{2}\sum_{i=1}^{N}\left(\frac{Y_i^2}{\rho_i^2}\right) - \frac{1}{2}\sum_{i=1}^{N}\log(\rho_i^2) - \frac{N}{2}\log(2\pi). \tag{3.3}$$

We must substitute into (3.3) a calculable quantity for $\rho_i^2$, hence we need such for $\sigma^2(t_{i-1})$ in (3.2). For this, we discretize the continuous-time volatility process, just as was done in Theorem 2.1. Thus, (2.4) reads, in the present notation,

$$\sigma_i^2 = \beta \Delta t_i + e^{-\eta \Delta t_i}\sigma_{i-1}^2 + \varphi e^{-\eta \Delta t_i}Y_i^2. \tag{3.4}$$

(3.4) is a GARCH-type recursion, so, after substituting $\sigma_{i-1}^2$ for $\sigma^2(t_{i-1})$ in (3.2), and the resulting modified $\rho_i^2$ in (3.3), we can think of (3.3) as the pseudo-log-likelihood function for fitting a GARCH model to the unequally spaced series.

The recursion in (3.4) is easily programmed and, taking as starting value for $\sigma^2(0)$ the stationary value $\beta/(\eta - \varphi)$, we can maximize the function $\mathcal{L}_N$ to get PMLEs of $(\beta, \eta, \varphi)$. In the next two sections, we apply this estimation approach to a data set of returns on the ASX200 share index and use simulation to study the properties of the estimates thus obtained. An alternative approach to estimating the COGARCH parameters based on the method of moments (MM) has been devised by [9]. Their results provide a baseline against which we can compare our procedure via simulations. By choosing suitable values for the simulation parameters, we are able to apply Theorem 2.1 in [9] to get moment estimates by their method as well and thus to compare their MM estimates with our PML estimates. However, the Haug *et al.* method only works for equally spaced series, so we have to restrict to this case to make the comparison.

### 3.1. Application to ASX stock index stock data

We used the PML method to fit a COGARCH model to a series consisting of 2529 log returns of the ASX200 stock index as listed on the Australian Stock Exchange taken once per trading day, March 1994 to March 2004. The data are shown in Figure 1.

Because of weekends and public holidays, the data are irregularly spaced, with the following frequencies of the inter-observation times:

| $\Delta t$ | 1 | 2 | 3 | 4 | 5 | 6 |
|---|---|---|---|---|---|---|
| frequency | 1991 | 13 | 483 | 24 | 17 | 1 |

For example, $\Delta t = 3$ corresponds to a regular weekend without additional public holidays. The data contain 2529 distinct values of the index returns, observed over a total time interval of $T = 3653$ days, and there are six distinct values of $\Delta t_i$. Simulations showed that instead of using equation (3.2) directly, one can use its first-order approximation, $\rho_i^2 = \sigma^2(t_{i-1})\Delta t_i$, without worsening the quality of the estimates.

The PML estimates were computed with an implementation of the Nelder–Mead optimization algorithm in C++. To avoid getting caught in a local (rather than the global) maximum of the pseudo-likelihood function, we used ten different starting simplices for each data set. The Nelder–Mead procedure was stopped when an accuracy of $10^{-14}$ in the location of the maximum of the function was reached. The approximate PMLEs were as follows ($\widehat{\beta}$ is multiplied by 365 to put it on an annualized basis, then the square root is taken so as to give a volatility rather than a variance estimate; approximate standard errors, calculated from the second derivative of $\mathcal{L}_N$ are in brackets):

$$\sqrt{365\widehat{\beta}} = 0.0237(0.0027); \qquad \widehat{\varphi} = 0.0685(0.0095); \qquad \widehat{\eta} = 0.0847(0.0085).$$

Note that our estimates satisfy the stationarity condition $\widehat{\eta} > \widehat{\varphi}$.

These estimates imply a long-run volatility value of $(365\widehat{\beta}/(\widehat{\eta} - \widehat{\varphi}))^{1/2} = 18.58\%$ p.a. By comparison, the *actual* standard deviation of the returns was 15.54% p.a. Estimates of the process $(\sigma^2(t))_{t \geq 0}$ at the observed time points can be calculated from

$$\widehat{\sigma}_i^2 = \widehat{\beta} + (1 - \widehat{\eta})\widehat{\sigma}_{i-1}^2 + \widehat{\varphi}(G(t_i) - G(t_{i-1}))^2$$

(cf. [9], equation (3.2)). Figure 2 shows the squared log returns for the first 1000 observations and, for comparison, the estimated annualized volatility.

To see how the volatility process evolves as the value of $\Delta t_i$ changes, we computed estimates of the transformed, rescaled, parameters $\omega_i := \beta(\Delta t_i)^2$, $\vartheta_i := \varphi e^{-\eta \Delta t_i} \Delta t_i$ and $\kappa_i := e^{-\eta \Delta t_i}$, which correspond to the discrete GARCH(1,1) parameterization. These are listed in columns 1–6 of Table 1 (but, again, we annualize the $\omega$ estimates and take the square root). Column 7 of Table 1 contains the GARCH(1,1) estimates obtained by treating the log returns as if they were equally spaced in time. As one would expect, treating the data as if they were equally spaced gives estimates corresponding approximately to a weighted averaging over the estimates in columns 1–6 of Table 1.

Quite commonly, financial analyses treat weekends or public holidays by assuming the data are contiguous over the missing period, thus, in effect, assuming that no information relevant to the market is transmitted on the missing days. This is not generally the case, of course, since, for example, trading in Australian stocks may be halted on a certain day on the ASX, while some or many of these stocks may nevertheless be traded on other international markets which are open at the time. While the corresponding information flow is probably not of the same strength as for a regular day's trading, we expect there will be some influence, and although our analysis above allows for unequally spaced time periods, it implicitly assumes that all data carry the same weight of information. But more generally, it can be argued that we should weight the observations in some way.

**Table 1.** Estimated parameters for various period lengths (columns 1–6) and GARCH estimates treating the data as equally spaced (column 7)

| $\Delta t_i$ | 1 | 2 | 3 | 4 | 5 | 6 | GARCH(1,1) |
|---|---|---|---|---|---|---|---|
| $(365\widehat{\omega}_i)^{1/2}$ | 0.0237 | 0.0473 | 0.0710 | 0.0946 | 0.1183 | 0.1419 | 0.0382 |
| $\widehat{\vartheta}_i$ | 0.0629 | 0.1157 | 0.1594 | 0.1953 | 0.2243 | 0.2472 | 0.0962 |
| $\widehat{\kappa}_i$ | 0.9188 | 0.8442 | 0.7756 | 0.7126 | 0.6548 | 0.6016 | 0.8434 |

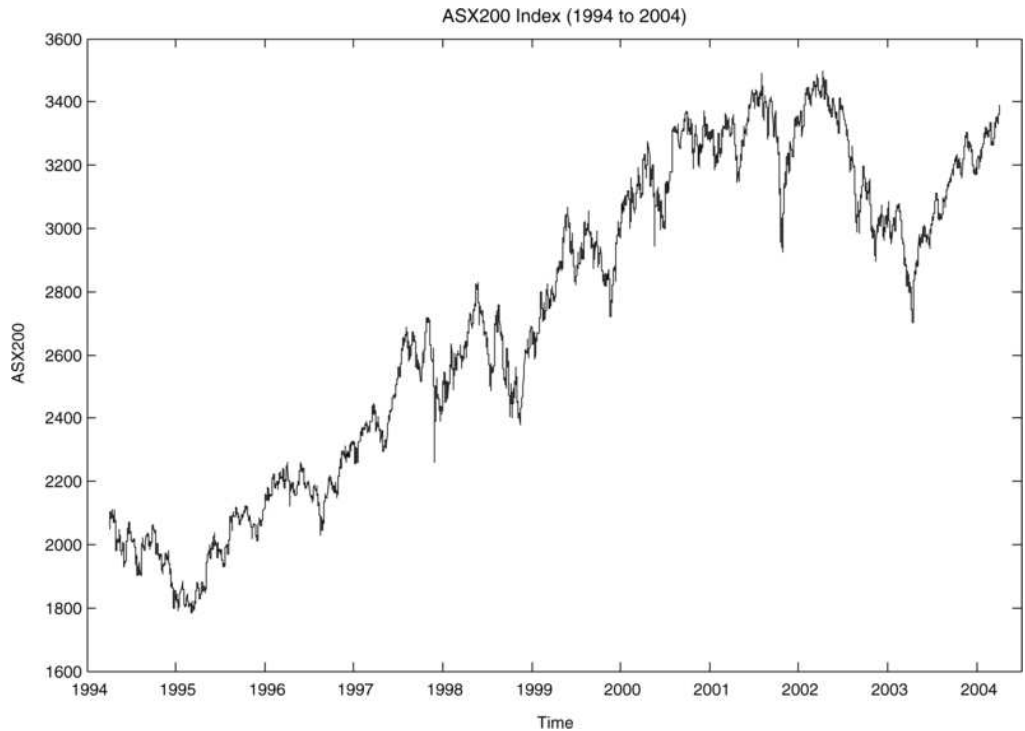

**Figure 1.** ASX200 stock index taken once per trading day, March 1994 to March 2004.

To investigate this, we extended the analysis using a function $w(\cdot)$ to weight the $\Delta t_i$, constraining the sum of the weights to remain the same as for the original analysis, that is,

$$\sum_{i=1}^{N} w(\Delta t_i) = \sum_{i=1}^{N} \Delta t_i = T. \tag{3.5}$$

In this setup, the function $w_0(\Delta t) := T/N$ represents an extreme case where the irregular spacing of the data is ignored, while the function $w_1 := \text{id}$ corresponds to our previous analysis where only the irregular spacing was taken into account. Another extreme case is to allow a separate parameter for each distinct value of $\Delta t$, rather than using the value of $\Delta t$ itself. For our data, this means fitting five extra parameters.

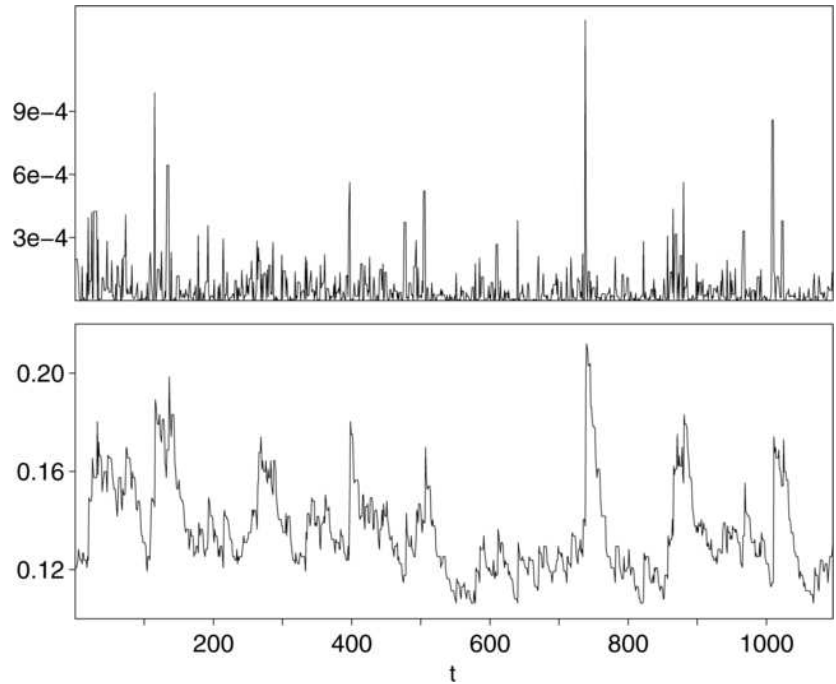

**Figure 2.** Top: squared log returns of ASX200 for the first three years (1096 days). Bottom: corresponding estimated annualized volatilities for the ASX index data.

Allowing the five extra parameters described gives a much better fit: the likelihood increases from 8649.61 for the original analysis to 8723.92. However, some of the extra parameters are very poorly determined (there is only one observation at $\Delta t = 6$, for example), and inspection of the parameter estimates suggested fitting the 1-parameter function $w_2(\Delta t) := \gamma \log(\Delta t) + c(\gamma)$, where $c$ is defined, depending only on $\gamma$, so that condition (3.5) is fulfilled. Replacing all $\Delta t_i$ by $w_2(\Delta t_i)$ and repeating the PML estimation, we find that the likelihood reduces only non-significantly from 8723.92 to 8721.00, still indicating a much better fit of the model to the data than the original unweighted model.

This application is by no means intended to be a sophisticated analysis of the ASX data set, which is beyond the scope of this paper. We use this irregularly spaced data example simply to illustrate the possibilities. Our main point is that the COGARCH model can be fitted directly to unequally spaced data exactly as it is, without the need to force it into an equally spaced setup in some way. Further, an approximation via the common GARCH(1, 1) model is easily adapted to the irregularly spaced case.

**Table 2.** Means over 1000 simulated estimates of $\beta$, $\varphi$ and $\eta$, average biases of the estimates, their mean absolute errors (MAE) and their root mean squared errors (RMSE) around the true value, for a time series of 5000 equally spaced observations from a COGARCH process driven by a compound Poisson process, using the PML method and the method of moments (MM)

| True | $\beta$ | | $\varphi$ | | $\eta$ | |
|---|---|---|---|---|---|---|
| | 1.0000 | | 0.0425 | | 0.0600 | |
| | PML | MM | PML | MM | PML | MM |
| mean | 1.2356 | 1.2487 | 0.0337 | 0.0448 | 0.0554 | 0.0672 |
| bias | 0.2356 | 0.2487 | −0.0088 | 0.0023 | −0.0046 | 0.0072 |
| MAE | 0.3799 | 0.4372 | 0.0099 | 0.0130 | 0.0125 | 0.0182 |
| RMSE | 0.5393 | 0.5892 | 0.0117 | 0.0146 | 0.0156 | 0.0231 |

### 3.2. Simulation study

In this section, our PML method is applied to simulated data sets, first with regularly spaced observations to allow a comparison with the results of [9], then with irregularly spaced data to see how much this influences the quality of the estimates.

For the first run, we simulated 1000 COGARCH data sets in which $T = 5000$ and observations occur at times $t = 1, 2, \ldots, 5000$. Thus, $N = 5000$, $\Delta t = 1$ and the ratio $T/N = 1$ approximates that of $T/N = 3653/2529 = 1.44$ in the ASX data. As driving Lévy process $L$, we chose a compound Poisson process with standard normal jump sizes and jump rate $\lambda = 1$. For the 'true' COGARCH parameters we took $\beta = 1$, $\varphi = 0.0425$ and $\eta = 0.06$. These values allow for the application of the method of moments to estimate the COGARCH parameters since all conditions of Theorem 2.1 in [9] are satisfied. To each data set, we applied the PML method to obtain estimates of $\beta$, $\varphi$ and $\eta$. In addition, we computed moment estimates using the method of [9]. The calculations were done in S-PLUS. Table 2 summarizes the results. It gives the mean over the 1000 simulated parameter estimates, the average bias of the estimates, their mean absolute error (MAE) and their root mean squared error (RMSE) around the true value, for both the PML and MM approaches. The standard errors of simulation of the values in the table are very small, as expected for a sample of size 1000, being less than 1%, for example, for the parameter estimates, so we do not report them.

The RMSE is a comprehensive error metric, combining the variance of the estimator around its true value, and its bias. Table 2 shows that our method reduces the RMSE of the moment estimates by 8.5% for $\beta$, by 20.0% for $\varphi$ and by as much as 32.5% for $\eta$. Note, however, that significant bias remains in all parameter estimates for both methods.

Next, we investigate how the quality of the PML estimates is affected when analyzing irregularly, rather than equally, spaced data. We simulated 1000 COGARCH processes according to the circumstances of the ASX data, thus, for 2659 values of $t$, occurring with the frequencies specified in Section 3.1 and with $\beta$, $\varphi$, and $\eta$ taken close to the PML estimates given there. Thus, we assume that we observed the COGARCH processes at exactly those times at which we observed the ASX data, encompassing a total time interval of $T = 3653$ days. Table 3 contains similar information as Table 2 for these simulations; in addition, in brackets in the last row are the relative RMSEs from the previous simulation study. These give some idea of how the quality of the estimates is affected by decreasing the number of observations from 5000 to 2529 and using irregularly, instead of equally, spaced data. In fact, we see that the quality of the estimates is not a great deal worse than from a data set with twice as many equally spaced observations.

## 4. Discussion

The GARCH methodology is now so well known and widely available that the model, or some variant of it, is fitted to economic or financial data almost as a matter of routine. One of the motivations for our present investigation, and that of Klüppelberg *et al.* in [14], in initiating the continuous-time model was with a view to applications such as the analysis of irregularly spaced time series, and options pricing.

Nelson's research [21] suggested that his limiting diffusion process, or some variant of it, might be useful as an assumed data generating process in a practical situation. (Leaving aside, at this point, considerations of the appropriateness of the model as a description of the data at hand, in which returns are probably not normally distributed, processes may have jumps, etc.) Assuming this, then, a very natural procedure is to consider fitting a GARCH model to (necessarily discrete) observations on the underlying process, then to substitute the resulting parameter estimates into a discrete option pricing algorithm (such as, e.g., the method of Ritchken and Trevor [23]), with the intention that the price thus obtained converges to the 'true' price, as it would be obtained from the underlying continuous-time model, when the mesh size of the approximation tends to 0.

**Table 3.** Means over 1000 simulated estimates of $\beta$, $\varphi$ and $\eta$, average biases of the estimates, their mean absolute errors (MAE) and their root mean squared errors (RMSE) around the true value, for a time series of length $T = 3653$ with 2529 irregularly spaced observations from a COGARCH process driven by a compound Poisson process, using the PML method. Last line: relative RMSE = RMSE/true parameter value and corresponding relative RMSEs from Table 2 (PML method) in brackets

|           | $\beta$         | $\varphi$        | $\eta$           |
|-----------|-----------------|------------------|------------------|
| True      | 1.5000          | 0.0690           | 0.0850           |
| mean      | 1.9573          | 0.0516           | 0.0718           |
| bias      | 0.4573          | $-0.0173$        | $-0.0132$        |
| MAE       | 0.6913          | 0.0197           | 0.0202           |
| RMSE      | 1.0100          | 0.0227           | 0.0242           |
| rel. RMSE | 0.6733 (0.5393) | 0.3291 (0.2744)  | 0.2848 (0.2606)  |

However, this plan goes awry at the first step because the potential nexus between the discrete GARCH estimation and the corresponding continuous-time parameters does not exist in a diffusion setting. This follows from Wang's [26] result, which shows that the GARCH estimates cannot identify the parameters in the continuous-time model, except in the degenerate case of the constant volatility model of Corradi [5]. This complication extends beyond options pricing methodologies, of course, but we stress that application because the discrete- to continuous-time step is transparent and crucial there.

In contrast, the COGARCH offers a class of models which appear as natural and appropriate analogs of the discrete GARCH models. The limit of our discrete-time GARCH approximating sequences is, in general, a jump process, not a diffusion and the close correspondence between the discrete- and continuous-time GARCH models makes it very plausible that they are statistically equivalent and will hence lead to consistent estimation. The evidence from the simulation study in Section 3.2 lends support to this conjecture. Nevertheless, it remains to be established, as do other large sample properties of the estimators and tests suggested by our approach. We leave this for the future.

## 5. Proofs

**Preliminaries**

Our pathwise construction relies on a 'first-jump' approximation to a Lévy process developed by Szimayer and Maller [25], which we present here in a general notation. Let $Z = \{Z(t): t \geq 0\}$ be a Lévy process with characteristic triplet $(\gamma^Z, 0, \Pi^Z)$, where $\gamma^Z$ relates to the standard truncation of $\Pi^Z$ in $[-1,1]$, and $Z(0) = 0$. Consider $Z$ on the compact interval $[0,T]$, which is divided into $N_n$ subintervals of length $\Delta t_i(n) := t_i(n) - t_{i-1}(n)$ for $i = 1, 2, \ldots, N_n$, where $0 = t_0(n) < t_1(n) < \cdots < t_{N_n}(n) = T$ is a deterministic partition of $[0,T]$ and $(N_n)_{n \geq 1}$ is a sequence of integers with $\lim_{n \to \infty} N_n = \infty$. Assume $\Delta t(n) := \max_{i=1,\ldots,N_n} \Delta t_i(n) \to 0$ as $n \to \infty$. Let $(m_n^Z)_{n \geq 1}$ be a positive sequence such that $\lim_{n \to \infty} m_n^Z = 0$ and define stopping times

$$\tau_{i,n}^Z = \inf\{t \in [t_{i-1}(n), t_i(n)) : |\Delta Z(t)| > m_n^Z\} \qquad \text{for } i = 1, \ldots, N_n, \qquad (5.1)$$

where $\Delta Z(t) := Z(t) - Z(t-)$. Define the 'first jump process' $(\overline{Z}_n(t) : 0 \leq t \leq T)$ by

$$\overline{Z}_n(t) = \sum_{i=1}^{N_n} \mathbf{1}_{\{\tau_{i,n}^Z \leq t\}} \Delta Z(\tau_{i,n}^Z) + t\left(\gamma^Z - \int_{m_n^Z \leq |z| \leq 1} z \Pi^Z(\mathrm{d}z)\right) \qquad \text{for } 0 \leq t \leq T. \quad (5.2)$$

The next proposition shows that, provided $\Delta t(n)$ and $m_n^Z$ converge to 0 at appropriate rates, the processes $\overline{Z}_n$ converge in probability to $Z$, uniformly for $t \in [0,T]$, as $n \to \infty$. Let $\overline{\Pi}^Z(z) = \Pi^Z\{[-z,z]^c\}$, $z > 0$, denote the tail of $\Pi^Z$ and assume $\overline{\Pi}^Z(0+) > 0$ to avoid trivialities.

**Proposition 5.1.** *Suppose* $\lim_{n\to\infty} \sqrt{\Delta t(n)}\,\overline{\Pi}^Z(m_n^Z) = 0$. *Then,* (i) *we have*

$$\sup_{0\leq t\leq T} |\overline{Z}_n(t) - Z(t)| \xrightarrow{\mathbb{P}} 0 \qquad as\ n\to\infty. \tag{5.3}$$

*If, in addition,* $\mathbb{E}|Z(1)| < \infty$ *and* $\mathbb{E}Z(1) = 0$, *we may replace* $\gamma^Z - \int_{m_n^Z \leq |z| \leq 1} z\Pi^Z(\mathrm{d}z)$ *by* $-\int_{|z|>m_n^Z} z\Pi^Z(\mathrm{d}z)$ *in* (5.2), *and* (5.3) *remains true.*

*If, further, we have* $\mathbb{E}(Z(1))^2 < \infty$, *then the convergence in* (5.3) *is, in fact, in* $\mathcal{L}_2$, *that is,* $\lim_{n\to\infty} \|\overline{Z}_n(t) - Z(t)\|_2 = 0$.

(ii) *If* $Z$ *is of finite variation with jump component* $Z^d(t) := \sum_{0<s\leq t} \Delta Z(s)$, *then*

$$\sup_{0\leq t\leq T} \left| \sum_{i=1}^{N_n} \mathbf{1}_{\{\tau_{i,n}^Z \leq t\}} \Delta Z(\tau_{i,n}^Z) - Z^d(t) \right| \xrightarrow{\mathbb{P}} 0 \qquad as\ n\to\infty. \tag{5.4}$$

**Proof.** (i) The claimed results follow immediately from Theorem 2.1 of [25]. The setup there is identical, except that the discretization of the state space they allow for is not needed here. So, in their theorem, we formally set $M(n) = \infty$ and $\Delta(n) = 0$, and identify $\overline{L}_n(t)$ in their notation with $\overline{Z}_n(t)$. Our equation (5.3) then follows from equation (2.11) of [25].

(ii) For Lévy processes of finite variation, truncation of the Lévy measure near 0 is not necessary and the truncation function $\mathbf{1}_{\{|z|\leq 1\}}$ can be dropped from the formulation. The same holds for the approximation scheme. Thus, (5.4) follows from (5.3).  □

**Proof of Theorem 2.1.** This proceeds in several steps. In parts (i)–(iii), the approximation procedures for $L(t)$, $\sigma^2(t)$ and $G(t)$ are outlined. The convergence, as stated in the theorem, is then shown in part (iv).

*Part (i): Approximation procedure for the underlying process* $L(t)$

The approximation procedure requires two stages. On one hand, we need a discrete GARCH approximating process satisfying (2.3) and (2.4). This does not come directly from the kind of approximation used in Proposition 5.1, but rather from the process $\widetilde{L}_n$ defined by

$$\widetilde{L}_n(t) := \sum_{i=1}^{N_n(t)} \sqrt{\Delta t_i(n)}\varepsilon_{i,n}, \qquad 0\leq t\leq T, n=1,2,\ldots. \tag{5.5}$$

Here, recall the $\varepsilon_{i,n}$ defined in (2.7). Recall, also, the first jump times $\tau_{i,n}$ defined in (2.5) and set $\tau_{i,n}^* = \tau_{i,n} \wedge t_i(n)$. Define the counting process

$$N_n(t) := \#\{i \in \mathbb{N} : \tau_{i,n}^* \leq t\}, \qquad 0 < t \leq T, \text{with } N_n(0) = 0.$$

$N_n(t)$ increases by 1 in each subinterval $(t_{i-1}(n), t_i(n)]$, $i = 1, 2, \ldots, n$, at the first time $\tau_{i,n}$ in the interval at which $L(t)$ changes in magnitude by more than $m_n$, or at $t_i(n)$, if there is no such change. Note that, finally, $N_n(t_{N_n(T)}(n)) = N_n(T) = N_n$.

As an intermediate step, we also need the sequence of processes defined by

$$\overline{L}_n(t) = \sum_{i=1}^{N_n(t)} \mathbf{1}_{\{\tau_{i,n}<\infty\}} \Delta L(\tau_{i,n}) - t \int_{|x|>m_n} x\Pi(\mathrm{d}x), \qquad 0 \le t \le T, \qquad (5.6)$$

to which we can apply Proposition 5.1. We have $\mathbb{E}|\overline{L}_n(1)| < \infty$, so, by Proposition 5.1, $\overline{L}_n$, as centered, converges in probability, uniformly on $[0,T]$, to $L$. Thus, to show that $\widetilde{L}_n \to L$ in probability, uniformly on $[0,T]$, we need only control the uniform distance of $\widetilde{L}_n$ from $\overline{L}_n$.

To estimate this, write $\overline{L}$ in terms of $\varepsilon_{i,n}$ as

$$\overline{L}_n(t) = \sum_{i=1}^{N_n(t)} (\varepsilon_{i,n}\xi_i(n) + \nu_i(n)) - t\int_{|x|>m_n} x\Pi(\mathrm{d}x).$$

Here,

$$\nu_i(n) := \mathbb{E}(\mathbf{1}_{\{\tau_{i,n}<\infty\}} \Delta L(\tau_{i,n})) = \frac{1 - \mathrm{e}^{-\Delta t_i(n)\overline{\Pi}(m_n)}}{\overline{\Pi}(m_n)} \int_{|x|>m_n} x\Pi(\mathrm{d}x)$$

and

$$\xi_i^2(n) := \mathrm{Var}(\mathbf{1}_{\{\tau_{i,n}<\infty\}} \Delta L(\tau_{i,n})) = \frac{1 - \mathrm{e}^{-\Delta t_i(n)\overline{\Pi}(m_n)}}{\overline{\Pi}(m_n)} \int_{|x|>m_n} x^2\Pi(\mathrm{d}x) - \nu_i^2(n)$$

are calculated from (2.6). Their asymptotic behaviors as $n \to \infty$ are

$$\max_{i=1,\ldots,N_n} \frac{|\nu_i(n)|}{\sqrt{\Delta t_i(n)}} \to 0 \quad \text{and} \quad \max_{i=1,\ldots,N_n} \left|\frac{\xi_i^2(n)}{\Delta t_i(n)} - 1\right| \to 0. \qquad (5.7)$$

To see this, use the inequality $1 - \mathrm{e}^{-x} \le x$, $x \ge 0$, and write

$$\frac{|\nu_i(n)|}{\sqrt{\Delta t_i(n)}} = O(\sqrt{\Delta t_i(n)}) \left|\int_{m_n<|x|\le 1} x\Pi(\mathrm{d}x) + \int_{|x|>1} x\Pi(\mathrm{d}x)\right|$$

$$\le O(\sqrt{\Delta t(n)})\left(\overline{\Pi}(m_n) + \left|\int_{|x|>1} x\Pi(\mathrm{d}x)\right|\right)$$

$$= O(\sqrt{\Delta t(n)\overline{\Pi}^2(m_n)}) + O(\sqrt{\Delta t(n)}).$$

Since $\lim_{n\to\infty} \Delta t(n)\overline{\Pi}^2(m_n) = 0$ by assumption, we get the result in (5.7) for $\nu_i(n)$, and then the result for $\xi_i^2(n)$ holds since $\int x^2\Pi(\mathrm{d}x) = \mathrm{var}(L(1)) = 1$ by assumption.

From (5.5) and (5.6), we have

$$\widetilde{L}_n(t) - \overline{L}_n(t) = \sum_{i=1}^{N_n(t)} (\sqrt{\Delta t_i(n)} - \xi_i(n))\varepsilon_{i,n} - \sum_{i=1}^{N_n(t)} \nu_i(n) + t\int_{|x|>m_n} x\Pi(\mathrm{d}x). \qquad (5.8)$$

Write $\sum_{i=1}^{N_n(t)} \Delta t_i(n) = t - r_n(t)$, where $0 \leq r_n(t) \leq \Delta t(n)$ a.s., and use the inequality $0 \leq x - 1 + e^{-x} \leq x^2/2$, $x \geq 0$, and the assumption that $\lim_{n \to \infty} \Delta t(n) \overline{\Pi}^2(m_n) = 0$ to get

$$\left| \sum_{i=1}^{N_n(t)} \nu_i(n) - t \int_{|x|>m_n} x\Pi(dx) \right|$$

$$= \left( \frac{1}{\overline{\Pi}(m_n)} \sum_{i=1}^{N_n(t)} (\Delta t_i(n)\overline{\Pi}(m_n) - 1 + e^{-\Delta t_i(n)\overline{\Pi}(m_n)}) + r_n(t) \right) \left| \int_{|x|>m_n} x\Pi(dx) \right|$$

$$\leq \left( O(\overline{\Pi}(m_n)) \sum_{i=1}^{N_n(t)} (\Delta t_i(n))^2 + \Delta t(n) \right) \left| \int_{|x|>m_n} x\Pi(dx) \right|$$

$$= (O(\sqrt{\Delta t(n)}) \sqrt{\Delta t(n)\overline{\Pi}^2(m_n)} + \Delta t(n)) \left| \int_{|x|>m_n} x\Pi(dx) \right|$$

$$= o(\sqrt{\Delta t(n)}) \left| \int_{|x|>m_n} x\Pi(dx) \right|.$$

Note that $\lim_{n \to \infty} \sqrt{\Delta t(n)} |\int_{|x|>m_n} x\Pi(dx)| = 0$ was shown in the proof of (5.7). Also, since the $(\varepsilon_{i,n})_{i=1,\ldots,N_n}$ are independent with means 0 and variances 1, and $\sum_{i=1}^{N_n(t)} \Delta t_i(n) \leq T$ the variance of the first term on the right-hand side of (5.8) is not larger than

$$\sum_{i=1}^{N_n(t)} |\sqrt{\Delta t_i(n)} - \xi_i(n)|^2 \leq T \max_{i=1,\ldots,N_n} \left| \frac{\xi_i(n)}{\sqrt{\Delta t_i(n)}} - 1 \right|^2 \to 0 \qquad \text{as } n \to \infty \qquad (5.9)$$

by (5.7). These arguments show that $\sup_{0 \leq t \leq T} |\overline{L}_n(t) - \widetilde{L}_n(t)| \xrightarrow{\mathbb{P}} 0$, as $n \to \infty$, as claimed, so we deduce from Proposition 5.1 the required convergence in probability, uniformly on $[0, T]$, of $\widetilde{L}_n$ to $L$.

*Part (ii): Approximation procedure for the variance process $\sigma^2(t)$*

Having defined the $\varepsilon_{i,n}$ in (2.7) and given the parameters $(\beta, \eta, \varphi)$, the variance process $\sigma_n^2$ is constructed using the GARCH(1, 1) equation (2.4). This can then be iterated (cf. [8, 21]) to get the explicit representation

$$\sigma_{i,n}^2 = \beta \sum_{j=1}^{i} \Delta t_j(n) \prod_{k=j+1}^{i} e^{-\eta \Delta t_k(n)} (1 + \varphi \Delta t_k(n) \varepsilon_{k,n}^2)$$

$$+ \sigma_{0,n}^2 \prod_{j=1}^{i} e^{-\eta \Delta t_j(n)} (1 + \varphi \Delta t_j(n) \varepsilon_{j,n}^2)$$

(5.10)

for $i = 0, 1, \ldots, N_n$ (take $\sum_{j=1}^{0} = 0$ and $\prod_{j=i+1}^{i} = 1$). Define a discrete-time process

$$X_{i,n} = \eta t_i(n) - \sum_{j=1}^{i} \log(1 + \varphi \Delta t_j(n) \varepsilon_{j,n}^2) \qquad \text{for } n = 1, 2, \ldots, \tag{5.11}$$

then define its continuous-time counterpart by interpolation:

$$\widetilde{X}_n(t) := X_{N_n(t), n} = \eta t_{N_n(t)}(n) - \sum_{i=1}^{N_n(t)} \log(1 + \varphi \Delta t_i(n) \varepsilon_{i,n}^2), \qquad 0 \le t \le T. \tag{5.12}$$

Note that $\widetilde{X}_n(\tau_{i,n}^*) = X_{i,n}$. Again, we wish to use the convergence result in Proposition 5.1, so we specify an auxiliary version of $\widetilde{X}$ as follows:

$$\overline{X}_n(t) = \eta t_{N_n(t)}(n) - \sum_{i=1}^{N_n(t)} \log(1 + \varphi \mathbf{1}_{\{\tau_{i,n} < \infty\}} (\Delta L(\tau_{i,n}))^2), \qquad 0 \le t \le T. \tag{5.13}$$

$\overline{X}$ is an approximation to $X$ as defined in (2.12), and $X$ is of finite variation, so, from Proposition 5.1, we have

$$\sup_{0 \le t \le T} |X(t) - \overline{X}_n(t)| \xrightarrow{\mathbb{P}} 0, \qquad \text{as } n \to \infty. \tag{5.14}$$

To check this, just compare (2.12) and (5.13), note that $\lim_{n \to \infty} t_{N_n(t)}(n) = t$ and set $Z(t) = X(t)$, $\overline{Z}(t) = \overline{X}(t)$ and $m_n^Z = \log(1 + \varphi m_n^2)$ in (5.1). Then $\tau_{i,n}^Z = \tau_{i,n}$ and part (ii) of Proposition 5.1 gives (5.14).

To establish the closeness of $\widetilde{X}_n$ to $\overline{X}$, write

$$|\log(1 + \varphi \Delta t_i(n) \varepsilon_{i,n}^2) - \log(1 + \varphi \mathbf{1}_{\{\tau_{i,n} < \infty\}} (\Delta L(\tau_{i,n}))^2)|/\varphi$$
$$\le |\Delta t_i(n) \varepsilon_{i,n}^2 - \mathbf{1}_{\{\tau_{i,n} < \infty\}} (\Delta L(\tau_{i,n}))^2| = |\Delta t_i(n) \varepsilon_{i,n}^2 - (\varepsilon_{i,n} \xi_i(n) + \nu_i(n))^2|$$
$$= |(\Delta t_i(n) - \xi_i^2(n)) \varepsilon_{i,n}^2 - 2 \xi_i(n) \varepsilon_{i,n} \nu_i(n) - \nu_i^2(n)|. \tag{5.15}$$

A similar argument as in (5.9) shows that the right-hand side of (5.15), when summed over $1 \le i \le N_n(t)$, tends in probability, uniformly on $[0, T]$, to 0. Thus, $\sup_{0 \le t \le T} |\overline{X}_n(t) - \widetilde{X}_n(t)| \xrightarrow{\mathbb{P}} 0$ as $n \to \infty$ and, using the triangle inequality, we conclude from (5.14) that

$$\sup_{0 \le t \le T} |X(t) - \widetilde{X}_n(t)| \xrightarrow{\mathbb{P}} 0 \qquad \text{as } n \to \infty. \tag{5.16}$$

Now we are in a position to show that an interpolated version of $\sigma_n^2$ approaches $\sigma^2(t)$, in the limit. Substituting in (5.10) for $X_{i,n}$ from (5.11), we can write, recalling $\sigma_{0,n}^2 = \sigma^2(0)$,

$$\sigma_{i,n}^2 = \beta e^{-X_{i,n}} \sum_{j=1}^{i} \Delta t_j(n) e^{X_{j,n}} + \sigma^2(0) e^{-X_{i,n}}. \tag{5.17}$$

Define the piecewise constant process

$$\widetilde{\sigma}_n^2(t) := \beta e^{-\widetilde{X}_n(t)} \sum_{i=1}^{N_n(t)} e^{\widetilde{X}_n(\tau_{i,n}^*)} \Delta t_i(n) + \sigma^2(0) e^{-\widetilde{X}_n(t)}, \qquad 0 \leq t \leq T. \tag{5.18}$$

Now, by (5.16), $e^{-\widetilde{X}_n}$ converges in probability, uniformly on $[0,T]$, to $e^{-X}$. To deal with the summation in (5.18), note that, except possibly for the last interval, where $i = N_n(t)$, we have $\widetilde{X}_n(\tau_{i,n}^*) = \widetilde{X}_n(t_i(n))$ since $\widetilde{X}_n$ can change value only at times $t = \tau_{i,n}^*$ and is constant elsewhere. Thus,

$$\sup_{0 \leq t \leq T} \left| \sum_{i=1}^{N_n(t)} \Delta t_i(n)(e^{\widetilde{X}_n(\tau_{i,n}^*)} - e^{\widetilde{X}_n(t_i(n))}) \right|$$

$$\leq 2\Delta t(n) \sup_{0 \leq t \leq T} e^{\widetilde{X}_n(t)} \leq 2e^{\eta T}\Delta t(n) \to 0$$

(note that $\widetilde{X}_n(t)$ is bounded above by $\eta T$, as is $X(t)$, for $0 \leq t \leq T$). Now, estimate

$$\sup_{0 \leq t \leq T} \left| \sum_{i=1}^{N_n(t)} \Delta t_i(n)(e^{X(t_i(n))} - e^{\widetilde{X}_n(t_i(n))}) \right|$$

$$\leq e^{X(t_i(n))} \sum_{i=1}^{N_n(T)} \Delta t_i(n) |1 - e^{\widetilde{X}_n(t_i(n)) - X(t_i(n))}|$$

$$\leq Te^{\eta T} \sup_{0 \leq s \leq T} |1 - e^{\widetilde{X}_n(s) - X(s)}|.$$

By (5.16), the last expression tends to 0 in probability. Finally, note that the discretely formed integral $\sum_1^{N_n(t)} e^{X(t_i(n))} \Delta t_i(n)$ converges in probability, uniformly on $[0,T]$, to the integral $\int_0^t e^{X(s)} \, ds$ by Theorem 21, Chapter II, of [22]. Hence, we deduce

$$\sup_{0 \leq t \leq T} \left| \sum_{i=1}^{N_n(t)} e^{\widetilde{X}_n(\tau_{i,n}^*)} \Delta t_i(n) - \int_0^t e^{X(s)} \, ds \right| \xrightarrow{\mathbb{P}} 0.$$

From (2.13) and (5.18), we now conclude that

$$\widetilde{\sigma}_n^2(t) \xrightarrow{\mathbb{P}} \beta e^{-X(t)} \int_0^t e^{X(s)} \, ds + \sigma^2(0) e^{-X(t)} = \sigma^2(t), \tag{5.19}$$

uniformly for $0 \leq t \leq T$.

*Part (iii): Approximation procedure for the COGARCH process $G(t)$*

In this section, we define a discrete integrated GARCH sequence $\widetilde{G}_n$ and prove its convergence to the continuous-time COGARCH process $G$. We take $G_{i,n}$ as in (2.3), thus

$$G_{i,n} = \sum_{j=1}^{i} \sigma_{j-1,n}\sqrt{\Delta t_j(n)}\varepsilon_{j,n}, \qquad i=1,\ldots,N_n,$$

with $\varepsilon_{j,n}$ and $\sigma_{j,n}^2$ satisfying (2.7) and (5.10). Interpolate to get a continuous-time version:

$$\widetilde{G}_n(t) = \sum_{i=1}^{N_n(t)} \sigma_{i-1,n}\sqrt{\Delta t_i(n)}\varepsilon_{i,n}, \qquad 0 \le t \le T. \tag{5.20}$$

By the definitions of $\widetilde{\sigma}_n$ and $\widetilde{L}_n$ in (5.5) and (5.18), we can write

$$\widetilde{G}_n(t) = \sum_{i=1}^{N_n(t)} \widetilde{\sigma}_n(\tau_{i-1,n}^*)\sqrt{\Delta t_i(n)}\varepsilon_{i,n} = \int_0^t \widetilde{\sigma}_n(s-)\,\mathrm{d}\widetilde{L}_n(s), \quad 0 \le t \le T,$$

so it is plausible that $\widetilde{G}_n(t) \xrightarrow{\mathbb{P}} G(t) = \int_0^t \sigma(s-)\,\mathrm{d}L(s)$, uniformly for $t \in [0,T]$. We confirm this as follows:

$$\widetilde{G}_n(t) = \sum_{i=1}^{N_n(t)} [\widetilde{\sigma}_n(\tau_{i-1,n}^*) - \sigma(\tau_{i-1,n}^*)]\sqrt{\Delta t_i(n)}\varepsilon_{i,n} + \sum_{i=1}^{N_n(t)} \sigma(\tau_{i-1,n}^*)D\widetilde{L}_n(\tau_{i,n}^*)$$

$$= \sum_{i=1}^{N_n(t)} [\widetilde{\sigma}_n(\tau_{i-1,n}^*) - \sigma(\tau_{i-1,n}^*)]\sqrt{\Delta t_i(n)}\varepsilon_{i,n}$$

$$+ \sum_{i=1}^{N_n(t)} \sigma(\tau_{i-1,n}^*)(D\widetilde{L}_n(\tau_{i,n}^*) - DL_n(\tau_{i,n}^*)) + \sum_{i=1}^{N_n(t)} \sigma(\tau_{i-1,n}^*)DL_n(\tau_{i,n}^*),$$

where $D\widetilde{L}_n(\tau_{i,n}^*) := \widetilde{L}_n(\tau_{i,n}^*) - \widetilde{L}_n(\tau_{i-1,n}^*)$ and $DL_n(\tau_{i,n}^*) := L_n(\tau_{i,n}^*) - L_n(\tau_{i-1,n}^*)$ for $i = 1, 2, \ldots, N_n$. Write the last expression as

$$\widetilde{G}_n(t) = M_{N_n(t),n} + Q_{N_n(t),n} + R_{N_n(t),n}, \tag{5.21}$$

where

$$M_{i,n} = \sum_{k=1}^{i}[\widetilde{\sigma}_n(\tau_{k-1,n}^*) - \sigma(\tau_{k-1,n}^*)]\sqrt{\Delta t_k(n)}\varepsilon_{k,n} = \sum_{k=1}^{i} a_{k-1,n}\sqrt{\Delta t_k(n)}\varepsilon_{k,n}$$

and

$$Q_{i,n} := \sum_{k=1}^{i}\sigma(\tau_{k-1,n}^*)(D\widetilde{L}_n(\tau_{k,n}^*) - DL_n(\tau_{k,n}^*)) = \sum_{k=1}^{i}\sigma(\tau_{k-1,n}^*)D_{k,n}, \qquad \text{say.}$$

First, we show that $M_{i,n}$ is uniformly asymptotically negligible. We plan to use Markov's and Doob's inequalities, but $\widetilde{\sigma}_n^2(t)$ does not necessarily have a finite expectation under our assumptions, so we need a truncation argument. For $v, C > 0$, write

$$\mathbb{P}\left(\max_{i=0,\ldots,N_n} |M_{i,n}| > v\right) \leq \mathbb{P}\left(\max_{i=0,\ldots,N_n} |M_{i,n}| > v, \sup_{0 \leq t \leq T} |\widetilde{\sigma}_n(t) - \sigma(t)| \leq C\right)$$
$$+ \mathbb{P}\left(\sup_{0 \leq t \leq T} |\widetilde{\sigma}_n(t) - \sigma(t)| > C\right).$$

The second term on the right tends to 0 as $n \to \infty$ by (5.19). The first term on the right is bounded by

$$\mathbb{P}\left(\max_{i=1,\ldots,N_n} \left|\sum_{k=1}^{i} a_{k-1,n} \mathbf{1}_{\{|a_{k-1,n}| \leq C\}} \sqrt{\Delta t_k(n)} \varepsilon_{k,n}\right| > v\right) = \mathbb{P}\left(\max_{i=1,\ldots,N_n} |M_{i,n}^C| > v\right), \quad \text{say.}$$

For each $n \geq 1$, $(M_{i,n}^C, \mathcal{F}_{\tau_{i,n}^*})_{i=0,\ldots,N_n}$ is a martingale. Use Markov's inequality and Doob's maximal quadratic inequality to obtain

$$\mathbb{P}\left(\max_{i=1,\ldots,N_n} |M_{i,n}^C| > v\right) \leq \frac{1}{v^2} \mathbb{E}\left(\max_{i=1,\ldots,N_n} (M_{i,n}^C)^2\right)$$
$$\leq \frac{4}{v^2} \mathbb{E}(M_{N_n,n}^C)^2 = \frac{4}{v^2} \sum_{k=1}^{N_n} \mathbb{E}(a_{k-1,n}^2 \mathbf{1}_{\{|a_{k-1,n}| \leq C\}}) \Delta t_k(n) \mathbb{E}(\varepsilon_{k,n}^2)$$
$$\leq \frac{4T}{v^2} \mathbb{E}\left(\min\left(\sup_{0 \leq t \leq T} |\widetilde{\sigma}_n(t) - \sigma(t)|^2, C^2\right)\right).$$

By (5.19) and the dominated convergence theorem, the above expression tends to 0 in probability. Hence, $\max_{i=1,\ldots,N_n} M_{i,n} \xrightarrow{\mathbb{P}} 0$ as $n \to \infty$.

Next, we deal with $Q_{i,n}$. From (2.12), we have $X(s) - X(t) \leq \sum_{s < u \leq t} \log(1 + \varphi(\Delta L(u))^2)$ when $0 \leq s < t$, so, from (2.13), we get

$$\mathbb{E}\left(\sup_{0 \leq t \leq T} \sigma^2(t)\right) \leq (\beta/\varphi + \mathbb{E}\sigma^2(0)) e^{\varphi T} =: C^*.$$

Further, for each $n \geq 1$, $(Q_{i,n}, \mathcal{F}_{\tau_{i,n}^*})_{i=0,\ldots,N_n}$ is a martingale, and we can use a similar argument as for $M_{i,n}$. Chebyshev's inequality and Doob's maximal quadratic inequality give

$$\mathbb{P}\left(\max_{i=1,\ldots,N_n} |Q_{i,n}| > v\right) \leq \frac{1}{v^2} \mathbb{E}\left[\max_{i=1,\ldots,N_n} \left(\sum_{k=1}^{i} \sigma(\tau_{k-1,n}^*) D_{k,n}\right)^2\right]$$
$$\leq \frac{4}{v^2} \mathbb{E}\left(\sum_{i=1}^{N_n} \sigma(\tau_{i-1,n}^*) D_{i,n}\right)^2$$

$$= \frac{4}{v^2} \sum_{i=1}^{N_n} \mathbb{E}(\sigma^2(\tau^*_{i-1,n})) \mathbb{E}(D^2_{i,n}).$$

An upper bound for this is

$$\frac{4}{v^2} \mathbb{E}\left(\sup_{0\leq t\leq T} \sigma^2(t)\right) \operatorname{Var}\left(\sum_{i=1}^{N_n} D_{i,n}\right) \leq \frac{4C^*}{v^2} \operatorname{Var}(\widetilde{L}_n(\tau^*_{N_n,n}) - L(\tau^*_{N_n,n})).$$

$$\leq \frac{4C^*}{v^2} \sup_{0\leq t\leq T} \mathbb{E}|\widetilde{L}_n(t) - L(t)|^2.$$

From (5.8), we can readily obtain that $\sup_{0\leq t\leq T} \mathbb{E}|\widetilde{L}_n(t) - \overline{L}(t)|^2 \to 0$ as $n \to \infty$. Also, from Proposition 5.1, $\sup_{0\leq t\leq T} \mathbb{E}|\overline{L}_n(t) - L(t)|^2 \to 0$ as $n \to \infty$. So, we have shown that the first and second summands in (5.21) are $o_P(1)$ as $n \to \infty$.

The third summand in (5.21), $R_{N_n(t),n}$, is a discrete stochastic integral with random partition $(\tau^*_{i,n})_{i=0,\ldots,N_n}$, where the mesh of the partition is bounded by $2\Delta t(n)$ and therefore tends to 0 a.s. Hence, Theorem 21, Chapter II, in [22] can be applied to show that this expression converges in probability, uniformly on $[0,T]$, to the stochastic integral $\int_0^\cdot \sigma(s-)\,\mathrm{d}L(s)$. So, finally,

$$\sup_{0\leq t\leq T} |\widetilde{G}_n(t) - G(t)| \xrightarrow{\mathbb{P}} 0, \qquad \text{as } n\to\infty. \tag{5.22}$$

*Part (iv): Convergence of the Skorokhod distance*

Finally, we have to transfer from the tilde processes $\widetilde{\sigma}^2_n(t)$ and $\widetilde{G}_n(t)$ to the desired approximating processes in (2.9). $\widetilde{\sigma}^2_n(t)$ and $\widetilde{G}_n(t)$ are constant between jump times $\tau^*_{i,n} = \tau_{i,n} \wedge t_i(n)$, for $n \geq 1$, so we can write for $0 \leq t \leq T$,

$$\sigma^2_n(t_i(n)) = \widetilde{\sigma}^2_n(\tau^*_{i,n}) \quad \text{and} \quad G_n(t_i(n)) = \widetilde{G}_n(\tau^*_{i,n}).$$

To obtain the convergence of $(G_n, \sigma^2_n)$ to $(G, \sigma^2)$ in the Skorokhod distance, it is crucial to note that both processes $\widetilde{\sigma}^2_n$ and $\widetilde{G}_n$ jump simultaneously and at most once in every interval $(t_{i-1}(n), t_i(n)]$ for $i = 1, \ldots, N_n$. The time change $\lambda(t)$ required in the Skorokhod distance can thus be specified pathwise as follows. On the grid $(t_i(n))_{i=1,\ldots,N_n-1}$, define

$$\lambda_n(t_i(n); \omega) = \tau^*_{i,n}(\omega) = \tau_{i,n}(\omega) \wedge t_i(n) \qquad \text{for } i = 1, \ldots, N_n - 1,$$

with $\lambda_n(0; \omega) = 0 = t_0(n)$ and $\lambda_n(T; \omega) = T = t_{N_n}(n)$, and interpolate piecewise linearly (hence continuously) between these points, thus obtaining a function $\lambda_n(\cdot; \omega)$ in $\Lambda$. By this construction, we see that

$$\sup_{0\leq t\leq T} |\lambda_n(t; \omega) - t| \leq \Delta t(n).$$

With the specification $\lambda_n(T;\omega) = T$ at the endpoint, required for $\lambda \in \Lambda$, we ignore any jump in the last subinterval $(t_{N_n-1}, T]$. However, the event $A_n = \{\tau_{N_n,n} \leq T\}$ has probability bounded by $\Delta t_{N_n}(n)\overline{\Pi}(m_n) = o(\sqrt{\Delta t(n)}) \to 0$ as $n \to \infty$, thus this modification is asymptotically negligible.

The definition of $\lambda_n(\cdot;\omega)$ allows us to write, on $A_n^C$,

$$\sigma_n^2(t) = \widetilde{\sigma}_n^2(\lambda_n(t;\omega)) \quad \text{and} \quad G_n(t) = \widetilde{G}_n(\lambda_n(t;\omega)) \qquad \text{for } 0 \leq t \leq T.$$

This implies

$$\sup_{0 \leq t \leq T} |\sigma_n^2(t) - \sigma^2(\lambda_n(t))| = \sup_{0 \leq t \leq T} |\widetilde{\sigma}_n^2(\lambda_n(t)) - \sigma^2(\lambda_n(t))|$$
$$= \sup_{0 \leq t \leq T} |\widetilde{\sigma}_n^2(t) - \sigma^2(t)|$$

and

$$\sup_{0 \leq t \leq T} |G_n(t) - G(\lambda_n(t))| = \sup_{0 \leq t \leq T} |\widetilde{G}_n(\lambda_n(t)) - G(\lambda_n(t))|$$
$$= \sup_{0 \leq t \leq T} |\widetilde{G}_n(t) - G(t)|.$$

Therefore, we can bound the Skorokhod distance by

$$\rho((G_n, \sigma_n), (G, \sigma^2)) \leq \sup_{0 \leq t \leq T} |\widetilde{G}_n(t) - G(t)| + \sup_{0 \leq t \leq T} |\widetilde{\sigma}_n^2(t) - \sigma^2(t)| + \Delta t(n)$$

and this expression tends to 0 in probability by (5.19) and (5.22), finishing the proof. $\square$

## Acknowledgements

This research was partially supported by ARC grant DP0664603. We are grateful to Professors T. Bollerslev and C. Klüppelberg, and two referees for constructive criticisms and helpful comments.